\documentclass[twocolumn,twoside,slac_two]{revtex4}
\usepackage{graphicx}
\usepackage{fancyhdr}
\newcommand{\kskl}{K_S^0 K_L^0}

\newcommand{\BR}{\mathcal{B}}

\newcommand{\ppbar}{p \overline{p}}

\newcommand{\psp}{\psi^{\prime}}
\newcommand{\pspto}{\psi^{\prime}\to}
\newcommand{\jpsi}{J/\psi}
\newcommand{\jpsito}{J/\psi\to}
\newcommand{\pspp}{\psi^{\prime\prime}}
\newcommand{\psppto}{\psi^{\prime\prime}\to}
\newcommand{\chicz}{\chi_{c0}}
\newcommand{\chico}{\chi_{c1}}
\newcommand{\chict}{\chi_{c2}}
\newcommand{\chicJ}{\chi_{cJ}}
\newcommand{\ppkk}{\pi^+\pi^-K^+K^-}
\newcommand{\EE}{e^+e^-}
\newcommand{\pipi}{\pi^+\pi^-}
\newcommand{\kskp}{K^0_S K^+ \pi^- + c.c.}

\newcommand{\pip}{\pi^+}
\newcommand{\pim}{\pi^-}
\newcommand{\piz}{\pi^0}
\newcommand{\ppp}{\pi^+\pi^-\pi^0}

\newcommand{\ddb}{D\bar{D}}

\newcommand{\rhopi}{\rho\pi}
\begin{document}

\title{Results from Charmonium Decays}

%

\author{Chang-Zheng Yuan}
\affiliation{Institute of High Energy Physics, Chinese Academy of
Sciences, Beijing, China}

\begin{abstract}

Recent results from BES and CLEOc experiments on charmonium decays
using $\jpsi$, $\psp$ and $\pspp$ data samples collected in $\EE$
annihilation are reviewed, including the measurement of the
relative phase between strong and electromagnetic decays of
$\psp$, the study of the ``$\rho\pi$ puzzle'' in $\jpsi$ and
$\psp$ decays, and the search for the non-$\ddb$ decays of
$\pspp$. The decays of $\chicJ$ produced in $\psp$ radiative
transition are also reviewed. These new results shed light on the
understanding of QCD.

\end{abstract}

\maketitle

\thispagestyle{fancy}


\section{Introduction}

BESII~\cite{besii} running at BEPC and CLEOc~\cite{cleo} running
at CESR are the two detectors operating in the $\tau$-charm energy
region. Both experiments have collected large data samples of
charmonium decays including 58~M $\jpsi$ events, 14~M $\psp$
events, and 33~pb$^{-1}$ data around $\pspp$ peak at BESII, and
4~M $\psp$ events, and 281~pb$^{-1}$ $\pspp$ events at CLEOc. To
study the continuum background in the charmonium decays, special
data samples at the center of mass energy lower than the $\psp$
mass were taken both at BESII ($\sqrt{s}=3.650$~GeV) and at CLEOc
($\sqrt{s}=3.671$~GeV), the luminosities are 6.4~pb$^{-1}$ and
21~pb$^{-1}$ respectively. These data samples are used for the
study of the hadron spectroscopy, the $D$ decay properties and the
CKM matrix, as well as the charmonium decay dynamics.

In this paper, we focus on the the extensive study of the
``$\rhopi$ puzzle'' in $\jpsi$ and $\psp$ decays, the relative
phase between strong and electromagnetic amplitudes of $\psp$
decays, and the non-$\ddb$ decays of $\pspp$. The study of
$\chicJ$ decays are also reviewed.

It should be noted that performance of the CLEOc detector is much
better than the BESII detector, especially in the photon
detection, this makes the 4~M $\psp$ events data sample at CLEOc
produces results with similar precision as from 14~M $\psp$ events
from BESII.

\section{Relative Phase in $\psp$ Decays}

It has been determined that for many two-body exclusive $\jpsi$
decays, like vector pseudoscalar (VP), vector vector (VV), and
pseudoscalar pseudoscalar (PP) meson decays and nucleon
anti-nucleon (N$\overline{\hbox{N}}$) decays, the relative phases
between the three-gluon and the one-photon annihilation amplitudes
are near $90^\circ$~\cite{suzuki,dm2exp,mk3exp,a00,a11,ann}.  For
$\psp$ decays, the available information about the phase is much
more limited because there are fewer experimental measurements. It
has been argued that the relative phases in $\psp$ decays should
be similar to those in $\jpsi$ decays~\cite{suzuki,gerard}, but
the analysis of $\psp$ to VP decays in Ref.~\cite{suzuki}
indicates this phase is likely to be around $180^\circ$. Another
analysis of this mode though shows the relative phase observed in
$\jpsi$ decays could also fit these decays~\cite{wymphase}, but it
could not rule out the $180^\circ$ possibility due to the big
uncertainties in the experimental data.

BES measured the branching fraction of $\pspto
\kskl$~\cite{besiipp} to be $(5.24\pm 0.47\pm 0.48)\times
10^{-5}$, together with the known branching fractions of $\pspto
\pi^+\pi^-$ and $\pspto K^+K^-$, two possible solutions of the
phase are found, which is either $-(82\pm 29)^\circ$ or $+(121\pm
27)^\circ$, following the procedure developed in
Ref.~\cite{wympp}. Benefits from the good detector performance,
CLEO measures all three pseudoscalar meson pair decay
modes~\cite{cleocpp}. The branching fraction of $\pspto \kskl$ is
$(5.8\pm 0.8\pm 0.4)\times 10^{-5}$, that of $\pspto K^+K^-$ is
$(6.3\pm 0.6\pm 0.3)\times 10^{-5}$; the signal of $\pspto
\pi^+\pi^-$ is not significant, and the upper limit of the
branching fraction is determined to be $2.1\times 10^{-5}$ at 90\%
C.L. CLEO measures the relative phase to be $(95\pm 15)^\circ$,
assuming there is no interference between continuum amplitude and
the resonance decay amplitudes, which is questionable since the
two components of the resonance decays has a relative phase around
$90^\circ$, as measured by both BESII and CLEOc experiments.
Following the same procedure developed in Ref.~\cite{wympp}, CLEO
measurements also result in two solutions for the phase, one at
around $-80$ degrees, while another at around $+120$ degrees, in
good agreement with the BES result. The error is large due to the
small statistics of the data samples, and there is no way to
determine which solution is the physical one.

\section{``$\rho\pi$ Puzzle'' in $\jpsi$ and $\psp$ Decay}

From perturbative QCD (pQCD), it is expected that both $\jpsi$ and
$\psp$ decaying into light hadrons are dominated by the
annihilation of $c\bar{c}$ into three gluons or one virtual
photon, with a width proportional to the square of the wave
function at the origin~\cite{appelquist}. This yields the pQCD
``12\% rule'',
\begin{equation}
 Q_h =\frac{{\cal B}_{\pspto h}}{{\cal
B}_{\jpsito h}} =\frac{{\cal B}_{\pspto \EE}}{{\cal B}_{\jpsito
\EE}} \approx 12\%.
\end{equation}
A large violation of this rule was first observed in decays to
$\rhopi$ and $K^{*+}K^-+c.c.$ by Mark~II~\cite{mk2}, known as {\it
the $\rhopi$ puzzle}, where only the upper limits on the branching
fractions were reported in $\psp$ decays. Since then, many
two-body decay modes of the $\psp$ have been measured by the BES
collaboration and recently by the CLEO collaboration; some decays
obey the rule while others violate
it~\cite{besres,cleocvp,psprhopi,besiipp,cleocpp}.

\subsection{$\psi\to \rho\pi$: Current Status}

The $\rho\pi$ mode of the vector charmonia decays is essential for
this study, since this is the first puzzling channel found in
$\jpsi$ and $\psp$ decays. The new measurements, together with the
old information, show us a new picture of the charmonium decay
dynamics.

\subsubsection{$\jpsi\to \ppp$}

BESII measured the $\jpsi\to \ppp$ branching fraction using its
$\jpsi$ and $\pspto \jpsi\pip\pim$ data samples~\cite{bes3pi}, and
BABAR measured the same branching fraction using $\jpsi$ events
produced by initial state radiative (ISR) events at
$\sqrt{s}=10.58$~GeV~\cite{babar3pi}. Together with the
measurement from Mark-II~\cite{mk2}, we get a weighted average of
$\BR(\jpsito\ppp)=(2.00\pm 0.09)\%$.

To extract the $\jpsito \rhopi$ branching fraction, partial wave
analysis (PWA) is needed to consider the possible contribution
from the excited $\rho$ states, the only available information on
the fraction of $\rhopi$ in $\jpsito \ppp$ was got from Mark-III.
Using the information given in Ref.~\cite{bill}, we estimate
$\frac{\BR(\jpsito \rhopi)}{\BR(\jpsito\ppp)}=1.17(1\pm 10\%)$,
with the error from an educated guess based on the information in
the paper since we have no access to the covariant matrix from the
fit showed in the paper. From this number and $\BR(\jpsito \ppp)$
got above, we estimate $\BR(\jpsito\rhopi)=(2.34\pm 0.26)\%$. This
is substantially larger than the world average listed by
PDG~\cite{pdg}, which is $(1.27\pm 0.09)\%$.

\subsubsection{$\psp\to \ppp$}

$\pspto \rhopi$ was studied both at BESII~\cite{psprhopi} and
CLEOc~\cite{cleocvp}. After selecting two charged pions and two
photons, clear $\piz$ signals are observed in the two photon
invariant mass spectra, the numbers of signals are found to be 229
and 196 from BESII and CLEOc respectively, and the branching
fraction of $\pspto \ppp$ is measured to be $(18.1\pm 1.8\pm
1.9)\times 10^{-5}$ and $(18.8^{+1.6}_{-1.5}\pm 1.9)\times
10^{-5}$ at BESII and CLEOc respectively. The two experiments give
results in good agreement with each other.

To extract the branching fraction of $\pspto \rhopi$, however,
BESII uses a PWA including the high mass $\rho$ states and the
interference between them, while CLEOc counts the number of events
by applying a $\rho$ mass cut. The branching fraction from BESII
is $(5.1\pm 0.7\pm 1.1)\times 10^{-5}$, while that from CLEOc is
$(2.4^{+0.8}_{-0.7}\pm 0.2)\times 10^{-5}$, the difference is
large. Although a big difference exists between the BESII and
CLEOc results, it does mean that the $\pspto \rhopi$ signal
exists, rather than the signal is completely missing as has been
guessed before. If we take a weighted average neglecting the
difference between the two measurements, we get \( \BR(\pspto
\rhopi)=(3.1\pm 0.7)\times 10^{-5} \).

Comparing $\BR(\pspto \rhopi)$ with $\BR(\jpsito \rhopi)$, one
gets
\[ Q_{\rhopi} = \frac{\BR(\pspto \rhopi)}{\BR(\jpsito \rhopi)} =
(0.13\pm 0.03)\%. \] The suppression compared to the 12\% rule is
obvious.

\subsection{Other Studies}

There are many more new measurements on $\psp$ decays for the
extensive study of the 12\%
rule~\cite{besres,bescleoc,cleocvp,besiipp,cleocpp}, among which
the VP modes are measured as first priority. The ratios of the
branching fractions are suppressed for almost all these VP modes
compared with the 12\% rule; while the PP modes, $\kskl$ and
$K^+K^-$, are enhanced compared with the 12\% rule. The
multi-hadron modes and the baryon-antibaryon modes are either
suppressed, or enhanced, or normal, which are very hard to be
categorized simply. The various models, developed for interpreting
specific mode can hardly find solution for all these newly
observed modes.

One observation is that many of the attempts to interpret the
$\rho\pi$ puzzle are based on the potential models for the
charmonium which were developed more than 20 years ago, as the
B-factories discovered many new charmonium states~\cite{xyz} which
are hard to be explained in the potential models, it may indicate
even the properties of $\jpsi$ and $\psp$ are not as expected from
the potential models. The further understanding of the other high
mass charmonium states may shed light on the understanding of the
low lying ones.

\section{Non-$\ddb$ Decays of $\pspp$}

Since $\pspp$ is above the $\ddb$ threshold, it decays
predominantly into charmed mesons, however, since the old
measurements may indicate big contribution of $\pspp$ charmless
decays~\cite{pdg}, and large fraction of charmless decays of
$\pspp$ is expected if $\pspp$ is a mixture of $S$- and $D$-wave
charmonium states and the mixing is responsible for the 12\% rule
violation in $\jpsi$ and $\psp$ decays~\cite{wympspp}, both BESII
and CLEOc experiments tried to search for the non-$\ddb$ decays of
$\pspp$ in both exclusive modes and inclusive measurement.

\subsection{Exclusive Decays}

It has been pointed out that the continuum amplitude plays an
important role in measuring $\pspp$ decays into light
hadrons~\cite{wympspp}. In fact, there are two issues which need
to be clarified in $\pspp$ decays, that is whether $\pspp$ decays
into light hadrons really exist, and if it exists, how large is
it. The searches for the non-$\ddb$ decays are performed by
comparing the cross sections on and off the $\pspp$ resonance
peak.

\subsubsection{$\pspp\to \ppp$}

By comparing the cross sections of $\EE\to \ppp$ at the $\pspp$
resonance peak ($\sqrt{s}=3.773$~GeV) and at a continuum energy
point ($\sqrt{s}=3.650$~GeV at BESII and $3.671$~GeV at CLEOc)
below the $\psp$ peak, both BESII and CLEOc found that
$\sigma(\EE\to \ppp)$ at continuum is larger than that at $\pspp$
resonance peak. The average of the two
experiments~\cite{psppbes,psppcleoc} are
    \[ \sigma(\EE\to \ppp)_{\hbox{on}}  =  7.5\pm 1.2~\hbox{pb}, \]
    \[ \sigma(\EE\to \ppp)_{\hbox{off}} = 13.7\pm 2.6~\hbox{pb}. \]
The difference, after considering the form factor variation
between 3.650 and 3.773~GeV, is still significant, and it
indicates that there is an amplitude from $\pspp$ decays which
interferes destructively with the continuum amplitude, and makes
the cross section at the $\pspp$ peak smaller than the pure
contribution of continuum process.

For the $\rhopi$ mode, BESII can only give upper limit of its
cross section due to the limited statistics of the data sample,
the upper limit at 90\% C. L. is found to be 6.0~pb~\cite{psppbes}
at the $\pspp$ peak, which is in consistent with the measurement
from CLEOc using a much larger data sample: \(\sigma(\EE\to
\rhopi)_{\hbox{on}}  = 4.4\pm 0.6~\hbox{pb} \)~\cite{psppcleoc};
while the cross section at the continuum is $8.0^{+1.7}_{-1.4} \pm
0.9$~pb measured by CLEOc.

\begin{figure}[htbp]
\begin{minipage}{7cm}
 \resizebox{14pc}{!}{\includegraphics[height=.5\textheight]{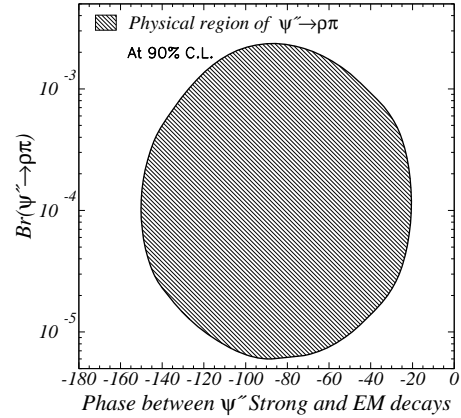}}
\end{minipage}
\caption{Physics region on $\BR(\psppto \rhopi)$ and the relative
phase ($\phi$) between $\pspp$ strong and electromagnetic decays
from BESII.} \label{besii_pspp}
\end{figure}

To extract the information on the $\psppto \rhopi$ branching
fraction, BESII developed a method based on the measured cross
sections at $\pspp$ resonance peak and at the
continuum~\cite{psppbes}. By neglecting the electromagnetic decay
amplitude of $\pspp$, there are two amplitudes which contribute to
the cross section at the $\pspp$ peak, the strong decay amplitude
of $\pspp$ and the continuum amplitude. Taking the continuum
amplitude as a real number, the $\pspp$ strong decay amplitude is
described as one real number for the magnitude, and one phase
between $\pspp$ strong and electromagnetic decays to describe the
relative phase between the two amplitudes. Since only two
measurements are available (at $\pspp$ peak and at continuum), one
can only extract $\pspp$ decay branching fraction as a function of
the relative phase. BESII measurement on the upper limit of the
$\EE\to\rhopi$ cross section at $\pspp$ peak, together with the
CLEOc measurement of the continuum cross section restrict the
physics region of the branching fraction and the relative phase as
shown in Fig.~\ref{besii_pspp}. From the Figure, we see that the
branching fraction of $\psppto \rhopi$ is restricted within
$6\times 10^{-6}$ and $2.4\times 10^{-3}$, and the phase is
between $-150^\circ$ and $-20^\circ$, at 90\% C.L.

The observation of the $\EE\to\rhopi$ signal at $\pspp$ peak and
the measurement of the cross section~\cite{psppcleoc} at CLEOc
further make the physical region in the branching ratio and
relative phase plane smaller: the CLEOc measurement gives a
similar outer bound of the physical region as BES gives, while
also indicates the central part of the physical region in
Fig.~\ref{besii_pspp} is not allowed by physics. By using a toy
Monte Carlo to simulate the CLEOc selection criteria and the
interference between the resonance and continuum amplitudes, we
found that the $\psppto \rhopi$ branching fraction could be either
$(2.1\pm 0.3)\times 10^{-3}$ or $(2.4^{+3.4}_{-2.0})\times
10^{-5}$ from the CLEOc measurements, if the relative phase
between $\pspp$ strong and electromagnetic decay amplitudes is
$-90^\circ$ as observed in $\jpsi$ and $\psp$
decays~\cite{wymphase}.

\subsubsection{Other Exclusive Modes}

The $\pspp$ decays into light hadrons were searched for in various
$\pspp$ decay modes, including two-bady and multi-hadron
modes~\cite{psppcleoc,psppmulti,besmulti}. However, only the
comparison between the cross sections at continuum and those at
$\pspp$ resonance peak are given, instead of giving the $\pspp$
decay branching fractions. In current circumstances, it is still
not clear whether the $\pspp$ decays into light hadrons with large
branching fractions, since, as has been shown in the $\rhopi$
case, there could be two solutions for the branching fraction, and
the two values could be very different.

As the luminosity at the $\pspp$ peak is large enough, current
study is limited by the low statistics at the continuum: at CLEOc,
the luminosity at continuum is more than an order of magnitude
smaller than that at peak, this prevents from a high precision
comparison between the cross sections at the two energy points.
One conclusion we can draw from the existing data is that the
measurements do not contradict with the assumption that the
relative phase between $\pspp$ strong and electromagnetic decay
amplitudes is around $-90^\circ$, and the $\pspp$ decays into
light hadrons could be large.

\subsection{Inclusive Measurements}

The inclusive non-$\ddb$ decays of $\pspp$ is searched for by
measuring the $\ddb$ cross section and the total hadronic cross
section above the $uds$ continuum contribution at the $\pspp$
resonance peak.

\begin{figure}[htbp]
\begin{minipage}{7cm}
 \resizebox{14pc}{!}{\includegraphics[height=.5\textheight]{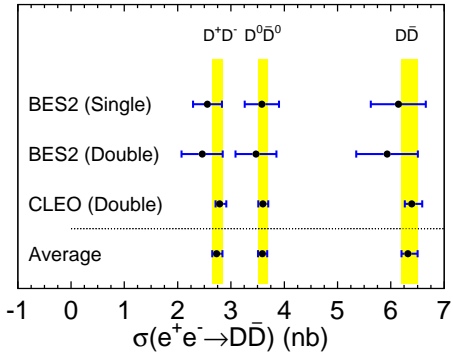}}
\end{minipage}
\caption{Measurements of the $\ddb$ cross section from BES and
CLEOc, and the weighted average of the measurements.} \label{ddb}
\end{figure}

BES and CLEO measure the $\ddb$ cross
section~\cite{besddb,cleocddb} using both single tag and double
tag methods, the results are shown in Fig.~\ref{ddb}. Good
agreement between BES and CLEO on the $D^+D^-$ and $D^0\bar{D^0}$,
as well as the total $\ddb$ cross sections are found. The weighted
average of the two experiments is $(6.32^{+0.18}_{-0.12})$~nb for
the total $\ddb$ cross section.

The total hadronic cross section of $\pspp$ decay is obtained by
comparing the cross sections at the $\pspp$ peak and the data
points below the resonance peak. The contributions of the
radiative tails of $\jpsi$ and $\psp$ are also subtracted. By
comparing this cross section and the $\ddb$ cross section, BES
found a non-$\ddb$ decay branching fraction of $(14.0\pm 1.7\pm
6.0)\%$~\cite{besmulti}, while CLEO measured the non-$\ddb$ cross
section of $(-0.01\pm
0.08^{+0.41}_{-0.30})$~nb~\cite{cleocnonddb}, which corresponds to
an upper limit of the non-$\ddb$ decay branching fraction of 11\%
at 90\% C. L.

Although the BES and CLEO measurements are not in contradiction
considering the large uncertainties, the big difference between
the two central values calls for a better understanding of the
procedure from which the total inclusive cross section is
obtained. An obvious difference between the two measurements is
how to consider the interference between the resonance and the
continuum amplitudes.

\section{$\chicJ$ Decays}

Since each $\chicJ$ ($J=0, 1, 2$) is produced about 10\% of the
$\psp$ decays, they are studied at BES to understand the P-wave
charmonium decay dynamics, as well as the light hadron
spectroscopy.

\subsection{$\chicJ\to VV$}

Two modes, $\omega\omega$~\cite{besww} and $\phi\phi$, were
measured recently, while the former is the first observation, the
latter improves the precision. The results are summarized in
Table~\ref{vv}, together with the measurement of $\chicJ\to
K^*(892)\bar{K}^*(892)$~\cite{kstarkstar}. These results are used
to predict the decay branching fractions of $\chicJ$ to other
vector meson pairs, like $\rho\rho$ and $\omega\phi$~\cite{zhaoq},
large double OZI suppressed amplitude is expected.

\begin{table}[htbp]
\begin{center}
\caption {Branching fractions of $\chicJ\to$ VV (in $10^{-3}$),
the results on $\phi\phi$ are BES preliminary.} \label{vv}
\begin{tabular}{|c|c|c|}\hline
   Decay mode  & $\chicz$   & $\chict$ \\\hline
$\omega\omega$ & $2.29\pm 0.58\pm 0.41$ & $1.77\pm 0.47\pm 0.36$
\\\hline
$\phi\phi$ & $0.94\pm 0.21\pm 0.14$  & $1.48\pm 0.26\pm 0.23$
\\\hline
$K^*(892)\bar{K}^*(892)$ & $1.78\pm 0.34\pm 0.34$  & $4.86\pm 0.56\pm 0.88$ \\
 \hline
\end {tabular}
\end{center}
\end {table}

\subsection{$\chicJ\to \Xi^-\bar{\Xi}^+$}

The importance of the Color Octet Mechanism (COM) for $\chicJ$
decays has been pointed out for many years~\cite{octet}, and
theoretical predictions of two-body exclusive decays have been
made based on it. Recently, new experimental results on $\chicJ$
exclusive decays have been reported~\cite{bes_chicj,bes}.  COM
predictions for many $\chicJ$ decays into meson pairs are in
agreement with experimental values, while predictions for some
decays into baryon pairs, for example, the branching fractions of
$\chicJ\to\Lambda \bar\Lambda$, disagree with measured values. For
further testing of the COM in the decays of the P-wave charmonia,
measurements of other baryon pair decays of $\chicJ$, such as
$\chicJ\to \Xi^-\bar{\Xi}^+$ is desired.

The measurement of $\chi_{c0}\to \Xi^-\bar{\Xi}^+$ is helpful for
understanding the Helicity Selection Rule (HSR)~\cite{brodsky},
which prohibits $\chi_{c0}$ decays into baryon antibaryon $(B\bar
B)$ pairs. However, the measured branching ratios for $\chi_{c0}$
decays into $\ppbar$ and $\Lambda \bar\Lambda$ do not vanish,
demonstrating a strong violation of HSR in charmonium decays.
Measurements of $\chi_{c0}$ decays into other baryon anti-baryon
pairs would provide additional tests of the HSR.

The measured branching fractions or upper limits are summarized in
Table~\ref{sumtab}~\cite{besxx}, along with some theoretical
predictions. Theoretically, the quark creation model (QCM)
predicts $B(\chi_{c0}\to \Xi^-\bar{\Xi}^+)=(2.3\pm 0.7)\times
10^{-4}$, which is consistent with the experimental value within
$1\sigma$. For $\chi_{c1}$ and $\chi_{c2}$ decays into
$\Xi^-\bar{\Xi}^+$, the measured upper limits cover both the COM
and QCM predictions. Within $1.8\sigma$ the branching fraction of
$\chi_{c0}\to \Xi^-\bar{\Xi}^+$ does not vanish. For further
testing of the violation of the HSR in this decay, higher accuracy
measurements are required.

\begin{table}[htbp]
\begin{center}
\caption{The comparison of the branching fractions or upper limits
for $\chicJ\to \Xi^-\bar{\Xi}^+$ between experimental values and
theoretical predictions. The COM predictions are from
Ref.~\cite{com_pre}, and the quark creation model (QCM)
predictions are from Ref.~\cite{qcm_pre}. The numbers are in unit
of $10^{-4}$, the upper limits are at 90\% C. L.} \label{sumtab}
\begin{tabular}{|c|c|c|c|}
\hline Channel & Branching Fraction &COM &QCM \\\hline $\chi_{c0}$
& $5.3\pm 2.7\pm 0.9$&--&$2.3\pm 0.7$  \\\hline &or
$<10.3$&&\\\hline $\chi_{c1}$ & $<3.4$ &$0.24$&--\\\hline
$\chi_{c2}$ & $<3.7$ &$0.34$ &$0.48\pm 0.21$\\\hline
\end{tabular}
\end{center}
\end{table}

\subsection{$\chicJ\to PPP$}

$\chicz$ decays into three pseudoscalars is forbidden by the
spin-parity conservation, while $\chict$ decays is suppressed due
to high orbital angular momentum. BES tries to measure the
branching fractions of $\chico$ decays into $\kskp$ and
$\eta\pipi$. Significant signals are observed and the branching
fractions are measured as
\[ \BR(\chico\to \kskp)=(4.1\pm 0.3\pm 0.7)\times 10^{-3}, \]
\[ \BR(\chico\to \eta \pipi)=(6.1\pm 0.8\pm 1.0)\times 10^{-3}. \]
The $\kskp$ events are mainly produced via $K^*(892)$ intermediate
state, and the $\eta\pipi$ events via $f_2(1270)\eta$ or
$a_0(980)\pi$. Except $\chico\to \kskp$, all other modes are first
observations. $\chict\to \kskp$ is also observed for the first
time with a branching fraction of $(0.8\pm 0.3\pm 0.2)\times
10^{-3}$. These results from BESII experiment are preliminary.

\subsection{$\chicz\to SS$}

Partial wave analysis of $\chi_{c0} \rightarrow \pi^+ \pi^- K^+
K^-$ is performed~\cite{beschic0} using $\chicz$ produced in
$\psp$ decays at BESII to study the pair production of scalars. In
14~M produced $\psp$ events, 1371 $\psi^{\prime} \rightarrow
\gamma \chi_{c0}$, $\chi_{c0} \rightarrow \pi^+\pi^-K^+K^-$
candidates are selected with around 3\% background contamination.

Besides $(\pi\pi)(KK)$ and $(K\pi)(K\pi)$ modes which are used to
study the scalars, $(K\pi\pi)K$ mode which leads to a measurement
of $K_1(1270)K$ and $K_1(1400)K$ decay processes is also included
in the fit. The PWA results are summarized in Table~\ref{pwachi}.
From these results, we notice that scalar resonances have larger
decay fractions compared to those of tensors, and such decays
provide a relatively clean laboratory to study the properties of
scalars, such as $f_0(980)$, $f_0(1370)$, $f_0(1710)$, and so
forth. The upper limits of the pair production of the scalar
mesons which are less significant are determined at the 90\% C.L.
to be at $10^{-4}$ level.

\begin{table}[htbp]
\begin{center}
\caption {Summary of the $\chicz\to \ppkk$ results, where $X$
represents the intermediate decay modes, and s.s. indicates signal
significance.} \label{pwachi}
\begin{tabular}{|c|c|c|}\hline
  Decay mode   & ${\cal B}[\chi_{c0}\rightarrow X \rightarrow$ & s.s. \\
  (X)&  $\pi^+\pi^- K^+K^-]$ $(10^{-4})$  & \\\hline
${f_0(980)f_0(980)}$&$3.46\pm
1.08^{+1.93}_{-1.57}$&$5.3\sigma$\\\hline
${f_0(980)f_0(2200)}$&$8.42\pm1.42^{+1.65}_{-2.29}$&$7.1\sigma$
\\\hline
${f_0(1370)f_0(1710)}$&$7.12\pm1.85^{+3.28}_{-1.68}$&$6.5\sigma$\\\hline
${K^*(892)^0\bar
K^*(892)^0}$&$8.09\pm1.69^{+2.29}_{-1.99}$&$7.1\sigma$\\\hline
${K^*_0(1430)\bar
K^*_0(1430)}$&$10.44\pm2.37^{+3.05}_{-1.90}$&$7.2\sigma$\\\hline
${K^*_0(1430)\bar K^*_2(1430)} +
c.c.$&$8.49\pm1.66^{+1.32}_{-1.99}$&$8.7\sigma$\\\hline
${K_1(1270)^{+}K^{-} + c.c.,}$&&\\\hline ~~~$K_1(1270)\to
K\rho(770)$&$9.32\pm1.83^{+1.81}_{-1.64}$&$8.6\sigma$\\\hline
${K_1(1400)^{+}K^{-} + c.c.,}$&&\\\hline ~~~$K_1(1400)\to
K^*(892)\pi$&$< 11.9$ (90\% C.L.) &$2.7\sigma$\\\hline
\end {tabular}
\end{center}
\end {table}

The above results supply important information on the
understanding of the natures of the scalar states~\cite{zhaoq}, as
well as the decay dynamics of $\chicJ$ decays into pair of scalar
particles.

\section{Summary}

There are many new results in charmonium decays from BESII and
CLEOc experiments. The decay properties of the vector charmonium
states have been studied for more than three decades, but they are
still far from being understood, one extreme example is the
``$\rhopi$ puzzle'' in $\jpsi$ and $\psp$ decays. Further studies
of all these are expected from the BESIII at BEPCII which will
start its data taking in late 2007.


\begin{acknowledgments}
I would like to thank the organizer for the invitation to give the
talk. I thank my colleagues in BES Collaboration for many helpful
discussions. This work was supported in part by National Natural
Science Foundation of China (10491303) and ``100-Talent Program''
of CAS under Contract No. U-25.
\end{acknowledgments}

\bigskip 

\end{document}